\documentclass[3p,sort&compress,times,twocolumn]{elsarticle}

\usepackage{lineno,hyperref}
\modulolinenumbers[2]

\journal{arXiv}

\usepackage{mathtools, cuted}
\usepackage{graphicx}   
\usepackage{amsmath}    
\usepackage{amssymb}    
\usepackage{threeparttable}
\usepackage{extarrows}
\usepackage{morefloats}
\usepackage[version=4]{mhchem} 
\usepackage{epstopdf} 
\usepackage[perpage]{footmisc}

\usepackage[final]{changes}
\DeclareGraphicsExtensions{.eps, .ps, .jpg, .pdf}
\graphicspath{{./}{./data/}{./figs/}}

\newcommand{\gl}{eq} 
\newcommand{\gls}{eqs}

\newcommand{\figname}{Figure~} 
\newcommand{\figsname}{Figures~}
\newcommand{\secname}{Section~}

\newcommand{\si}{SI}
\usepackage{mfirstuc}
\MFUnocap{in}
\MFUnocap{of}
\MFUnocap{that}
\MFUnocap{by}
\MFUnocap{to}
\MFUnocap{on}
\MFUnocap{and}
\MFUnocap{in}
\MFUnocap{the}
\MFUnocap{for}
\MFUnocap{with}
\begin{document}

\begin{frontmatter}

\title 
{Self-assembled monolayers of oligophenylenes stiffer than steel and silicon, possibly even stiffer than \ce{Si3N4}}

\author{Ioan B\^aldea}
\address{Theoretical Chemistry, Heidelberg University, Im Neuenheimer Feld 229, D-69120 Heidelberg, Germany}
\fntext[myfootnote]{ioan.baldea@pci.uni-heidelberg.de}
\begin{abstract}
  To quantify charge transport through molecular junctions fabricated using
  the conducting probe atomic force microscopy (CP-AFM) platform, information on the number
  of molecules $N$ per junction is absolutely necessary. $N$ can be currently obtained only
  via contact mechanics, and the Young's modulus $E$ of the self-assembled monolayer (SAM) 
  utilized in the key quantity for this approach.
  The experimental determination of $E$ for SAMs of CP-AFM junctions
  fabricated using oligophenylene dithiols (OPDn, $1 \leq n \leq 4$)
  and gold electrodes turned out to be too challenging.
  Recent measurements (Z.~Xie et al, J.~Am.~Chem.~Soc.~139 (2017) 5696) 
  merely succeeded to provide a low bound estimate ($E \approx 58$\,GPa). 
  It is this state of affairs that motivated the present theoretical investigation.
  Our microscopic calculations yield values $E \approx 240 \pm 6$\,GPa for the OPDn SAMs
  of the aforementioned experimental study, which are larger than those of steel ($ E \approx 180 - 200$\,GPa)
  and silicon ($E \approx 130 - 185$\,GPa). 
  The fact that the presently computed $E$ is much larger than the aforementioned
  experimental lower bound explain why experimentally measuring $E$ of OPDn SAM's is so challenging.
  Having $E \approx 337 \pm 8$\,GPa,
  OPDn SAMs with herringbone arrangement adsorbed on fcc (111)Au are even stiffer than \ce{Si3N4} ($ E \approx 160 - 290$\,GPa).
\end{abstract}


\end{frontmatter}

\section{Introduction}
\label{sec:intro}
Among the various platforms to fabricate molecular junctions
\cite{Reed:97,Tao:03,Loertscher:07,Venkataraman:06,McCreery:09,Ruitenbeek:09,Reed:09,Reed:11,CuevasScheer:17,Baldea:2015e},
conducting probe atomic force microscopy (CP-AFM) 
\cite{Frisbie:00,Frisbie:01,Frisbie:02,Frisbie:02b,Frisbie:03,Frisbie:05a,Beebe:06,Beebe:08,Frisbie:06,Boer:08,Tan:10,Frisbie:11,Frisbie:11b}
is an approach pioneered by Frisbie's group \cite{Frisbie:00} that offers a series of advantages \cite{Baldea:2017e}.
CP-AFM junctions consist of bundles containing a number $N$ of molecules
trapped between the metal-coated cantilever (AFM tip) and substrate covered by
a self-assembled monolayer (SAM). 
Comparison between CP-AFM-based transport properties and those measured, e.g.,
for single molecule STM (scanning tunneling microscopy) break junctions or fabricated using crossed wire techniques
\cite{Kushmerick:02,Kushmerick:02b,Kushmerick:05,Beebe:06,Beebe:08}
involves the knowledge of the number of molecules $N$,
a quantity that, obviously, plays a crucial role from this perspective.
Given the impossibility of a direct determination from experiment, combining
contact mechanics \cite{Johnson:85,Butt:05,Haugstad:12}
to obtain the contact area $A$
with Rutherford backscattering (RBS) and/or nuclear reaction analysis (NRA)
to measure the surface coverage $\Sigma$ 
is the state-of-the-art approach
\cite{Baldea:2015d,Frisbie:16e,Baldea:2017e,Baldea:2018c,Baldea:2019a,Baldea:2019d,Baldea:2019h}
to estimate $N = \Sigma A$. 

While the aforementioned nuclear methods enable the direct
determination of the surface coverage $\Sigma$,
models developed in contact mechanics pose certain problems to reliably
estimate the contact area $A$
between the AFM tip and the self-assembled monolayer (SAMs) under investigation.
To deduce $A$, various 
models were developed in contact mechanics \cite{Johnson:71,Derjaguin:75,Dugdale:60,Maugis:92,Butt:05,Haugstad:12}.
To apply these models, SAM's Young modulus of elasticity $E$ is a key quantity needed.
Unfortunately, as noted earlier \cite{Baldea:2017e} and elaborated in \secname\ref{sec:method},
the experimental determination of $E$ is problematic. 

So, the best one can do at present is to theoretically compute $E$ by investigating
nanoelastic properties of the molecules of interest subject to compressive or tensile forces.
It is this state of affairs that motivated the present theoretical investigation
on molecules \ce{OPDn\equiv HS\bond{1}(C6H4)_n\bond{1}SH} ($1\leq n \leq 4$) of the benchmark
oligophenylene dithiol family
\cite{Reed:97,Guo:11,Reddy:11b,Reddy:12b,Tao:13,Baldea:2015d,Baldea:2017e}.

While primarily aiming at providing values of $E$ to be subsequently utilized in
ongoing molecular electronics studies,
the results reported below also provide an explanation why a direct
experimental determination of Young's modulus of SAMs based on
aromatic oligophenylene molecules is currently too challenging. We found that OPDn SAMs are much stiffer
than the lower bound estimate ($ E \approx 58$\,GPa) deduced in recent experiments \cite{Baldea:2017e}
may suggest. They are stiffer than steel and silicon, possibly also stiffer than
\ce{Si3N4}.
\section{Method}
\label{sec:method}
\begin{figure}
  \centerline{\includegraphics[width=0.45\textwidth]{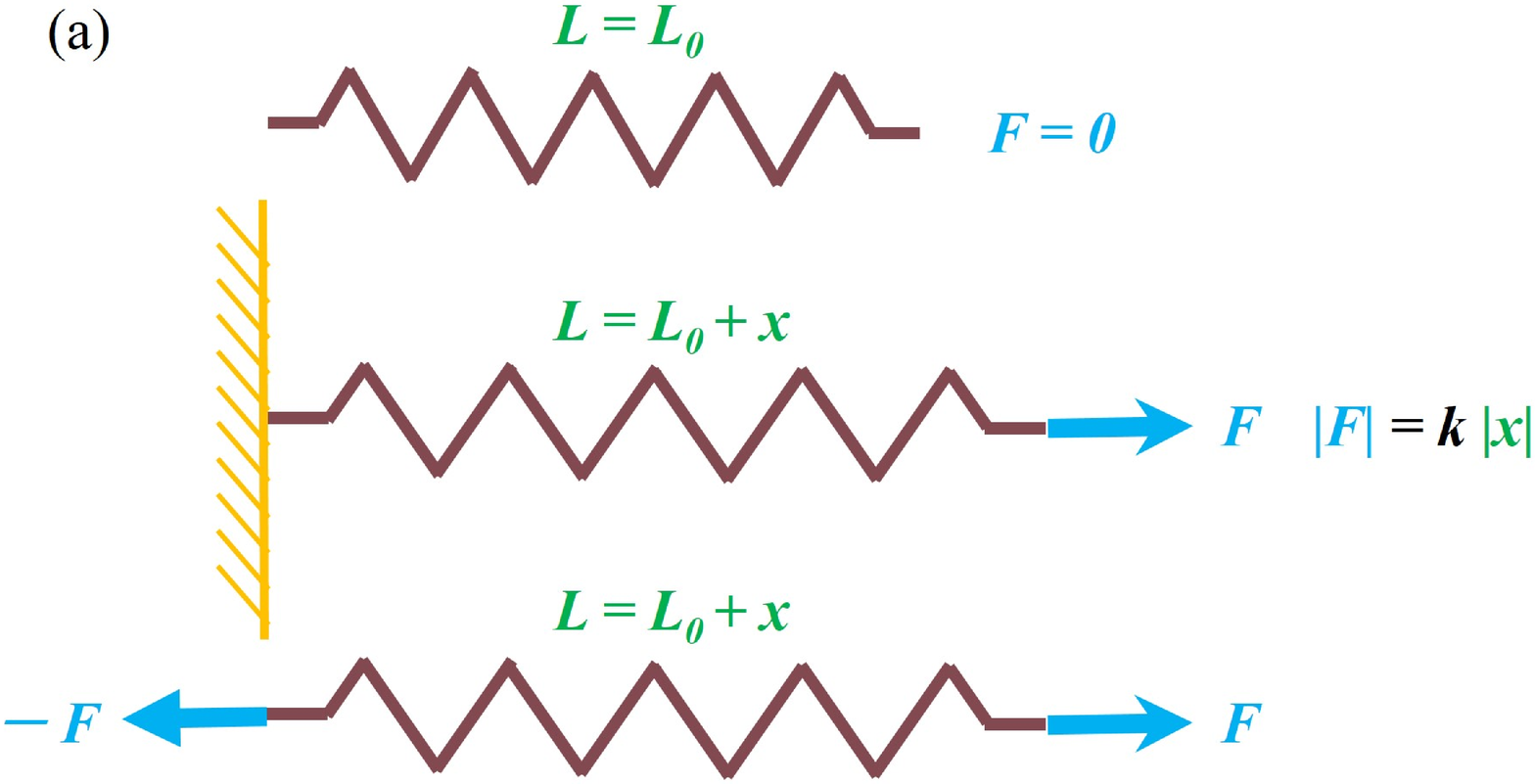}}
$ $\\[5ex]
  \centerline{\includegraphics[width=0.45\textwidth]{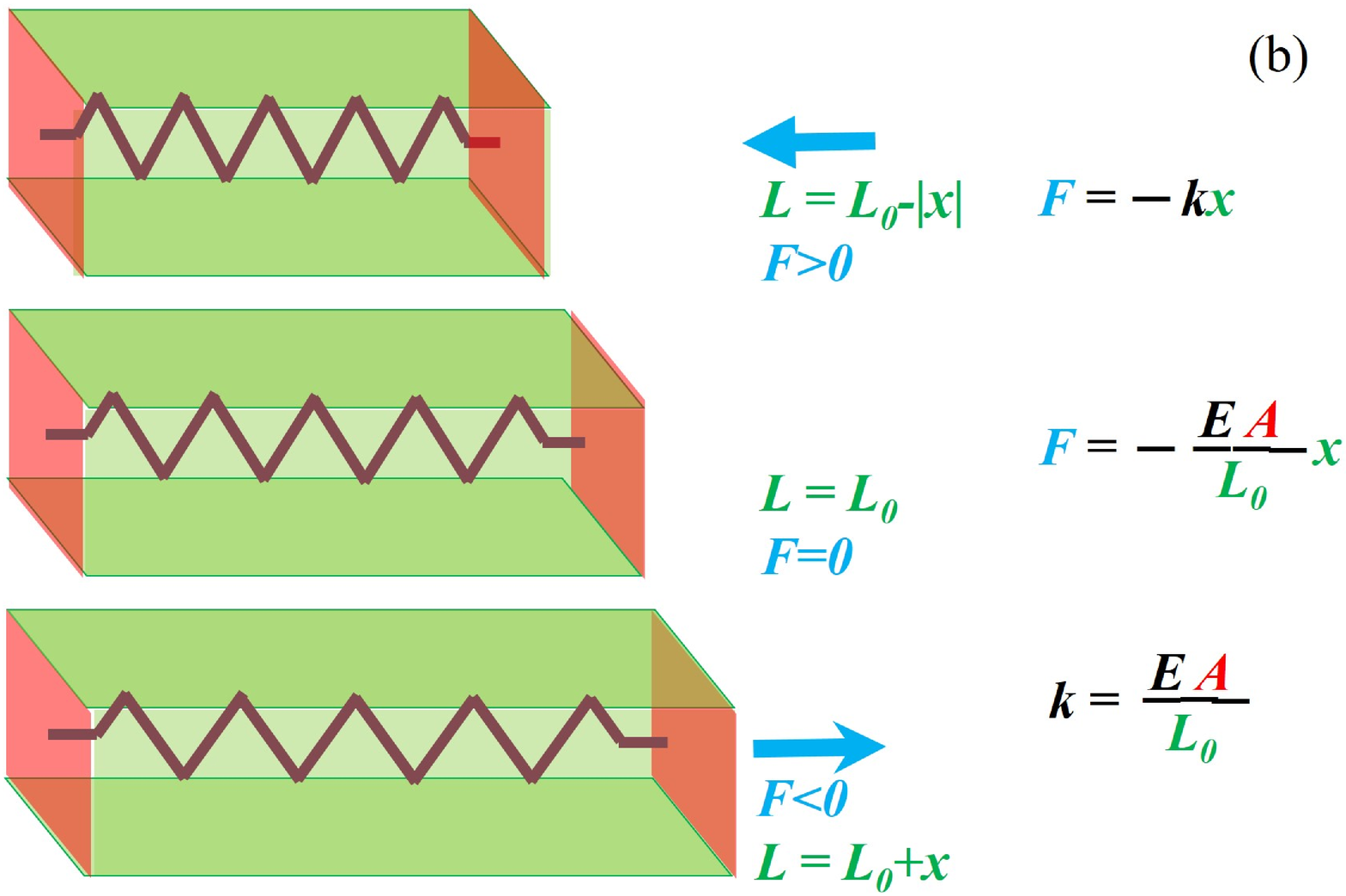}}
$ $\\[5ex]
  \centerline{ \includegraphics[width=0.45\textwidth]{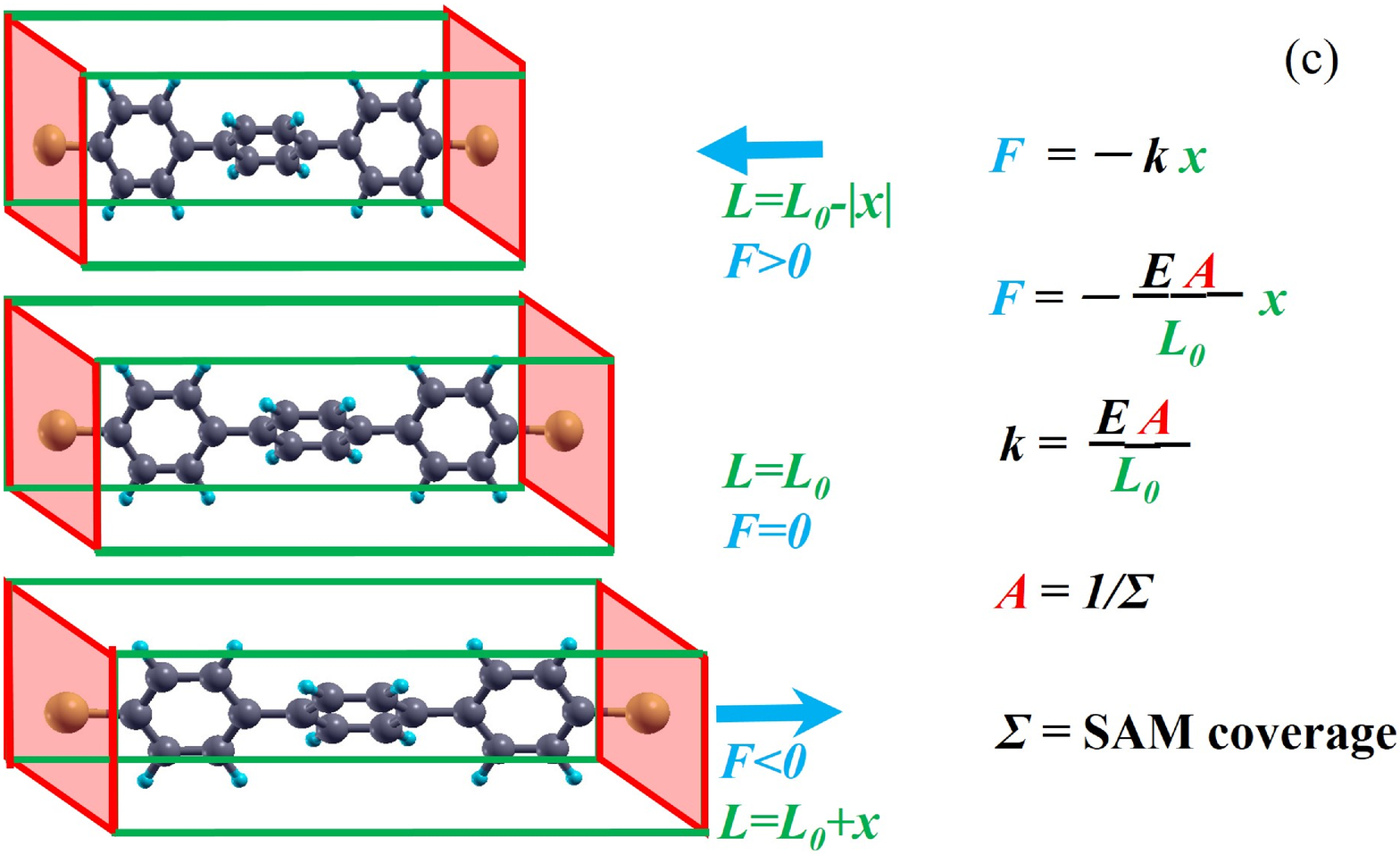}}
\caption{Schematic representation of the method to compute the elastic constant of a molecule.} 
\label{fig:schematic}
\end{figure}

Quantum chemical computations accomplished in conjunction with the present
investigation used the GAUSSIAN 16 package \cite{g16} 
on the bwHPC
platform. We performed geometry optimizations for
molecules subject to axial mechanical forces (cf.~\figname\ref{fig:schematic}).
These calculations were based on the density functional theory using the Becke
three-parameter B3LYP hybrid exchange-correlation functional
\cite{Becke:88,Becke:93a}. For consistency with our recent 
\cite{Baldea:2020a,Baldea:2020b,Baldea:2020c,Baldea:2020h}
and ongoing works on related systems,
we used 6-311++G(D, P) basis sets \citep{Petersson:88,Petersson:91}
although including diffuse basis functions is in fact necessarily required
only at larger molecular elongations beyond the linear (elastic) deformation
regime investigated in this paper.
\section{Results and Discussion}
\label{sec:results}
\subsection{Preliminary Remarks}
\label{sec:remarks}
As anticipated in Introduction,
SAM's modulus of elasticity $E \equiv E_s$ represents an important challenge for experimentalists. 
Insufficient signal-to-noise ratio makes it impossible to
reliable estimate an effective Young modulus for OPDn/Au from
tiny differences
in indentation depths measured with (OPDn/Au) and without (bare Au)
SAM adsorbed on gold (C.~D.~Frisbie and Z.~Xie, private communication).
To this, from a more general perspective, one should still add the aggravating point 
that the exact determination of $E$
merely from force-distance curves is impossible (see ref.~\cite{Butt:05}, page 41 for details):
the mutual dependence between the slope of the contact line and the jump-off-contact is expressed in terms
of a parameter ($\lambda $ in ref.~\cite{Baldea:2017e}) 
for which (among other nontrivial things)
information on SAM's modulus of elasticity $E$ 
is required; 
see, e.g., ref.~\cite{Butt:05}, page 41 for specific details.
To avoid misunderstandings, one should emphasize here that the aforementioned $E = E_s$ as used in
contact mechanics calculations is 
a property of the SAM, which does not include the
elastic interaction with the AFM tip. 
Contact mechanics does account for the tip-SAM interaction (as it should do), but it does it 
via the work of adhesion ($\gamma$ in ref.~\cite{Baldea:2017e}). The so-called ``effective'' Young's modulus entering
formulas of the various contact mechanics models \cite{Johnson:71,Derjaguin:75,Dugdale:60,Maugis:92,Butt:05,Haugstad:12}
depends both on SAM's ($E \equiv E_s$) and tip's ($E_t$) moduli of elasticity; still these 
are ``intrinsic'' Young's moduli, which
refer to tip and SAM \emph{isolated}
(i.e.~elastically decoupled) of each other.

Having said this, and given the fact that elastic properties of metals (gold in ref.~\cite{Baldea:2017e})
used for (coating) AFM tips are known, what we have to do is to focus on the elastic properties of
monolayers of OPDn molecules which, albeit not (mechanically) coupled to the AFM tip,
still have the same spatial structure as the real SAM.

In this context, one should note that OPDn molecules in SAMs utilized to fabricate the recently investigated
CP-AFM junctions \cite{Baldea:2015d,Baldea:2017e}
do not resemble to those of ordinary (organic) solids or liquids. Both (ellipsometry and XPS
\cite{Baldea:2015d,Baldea:2017e,Baldea:2019d})
experiments and theory \cite{Baldea:2017f,Baldea:2019b,Baldea:2019f}
found that OPDn molecules stand nearly vertical on metal.
The average intermolecular spacing 
deduced from the coverage ($\Sigma \simeq 3.3\,\mbox{molec/nm}^2$ \cite{Baldea:2017e})
amounts to $\sim 5.5$\,{\AA}. So, we are dealing with monolayers wherein parallel
OPDn molecules are sufficiently apart of each other, and considering elastic properties of
strands of OPDn molecules \emph{isolated} of each other is a legitimate description.
Parenthetically, this one dimensional picture in SAM's transverse direction
is additionally supported by other pieces of recent experimental
evidence \cite{Baldea:2015d,Baldea:2017e} revealing a
charge transport through individual OPDn mo\emph{isolated} \emph{isolated} molecules weakly interacting
among themselves in transverse direction.

The foregoing analysis made it clear that what we ultimately need to compute are elastic properties of isolated
OPDn molecules. 
\subsection{Elastic Properties of OPDn and OPn Molecules}
\label{sec:kappa-K-E}
Constrained optimization imposing a fixed values of the
distance $L(\ce{X_1, X_2}) \to L = L_0 + x$ between the two
(non-hydrogen) \ce{X_{1,2}} atoms at the two molecular ends
(\ce{X_{1,2}=S} for OPDn and \ce{X_{1,2}=C} for OPn) 
larger or smaller than the equilibrium
value ($L_0$) straightforwardly allows the determination of the
tensile or compressive forces
$F = F(x)$ (of opposite direction and equal magnitude) 
exerted on the \ce{X_1} and \ce{X_2} end atoms.

\begin{figure*}
  \centerline{\includegraphics[width=0.45\textwidth]{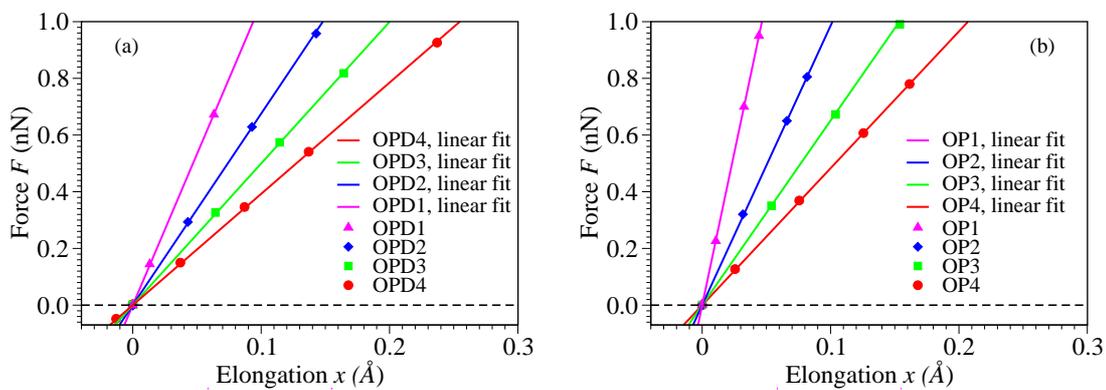}
    \includegraphics[width=0.45\textwidth]{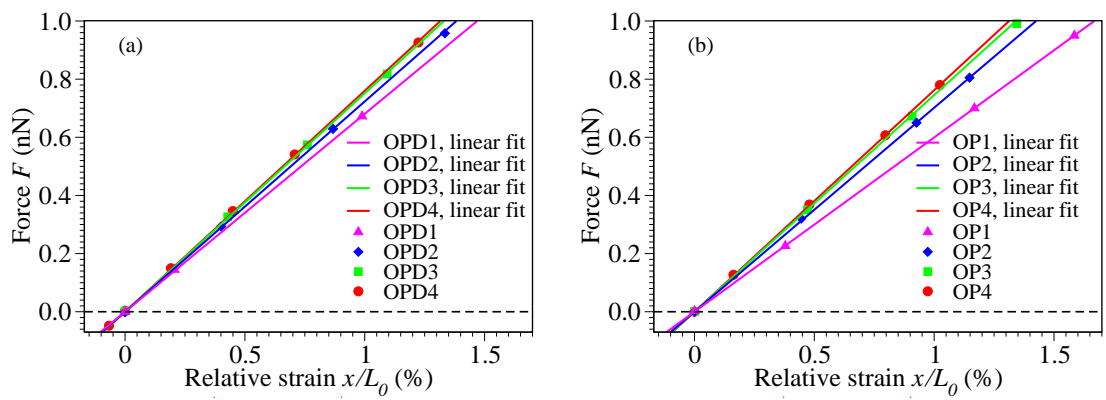}}
\caption{Results for the mechanical (here, $F>0$, tensile or compressing $F<0$) force
                 as a function of molecular length $L$: (a) OPDn and (b) OPn.} 
\label{fig:F-vs-x}
\end{figure*}

\begin{figure*}
  \centerline{\includegraphics[width=0.45\textwidth]{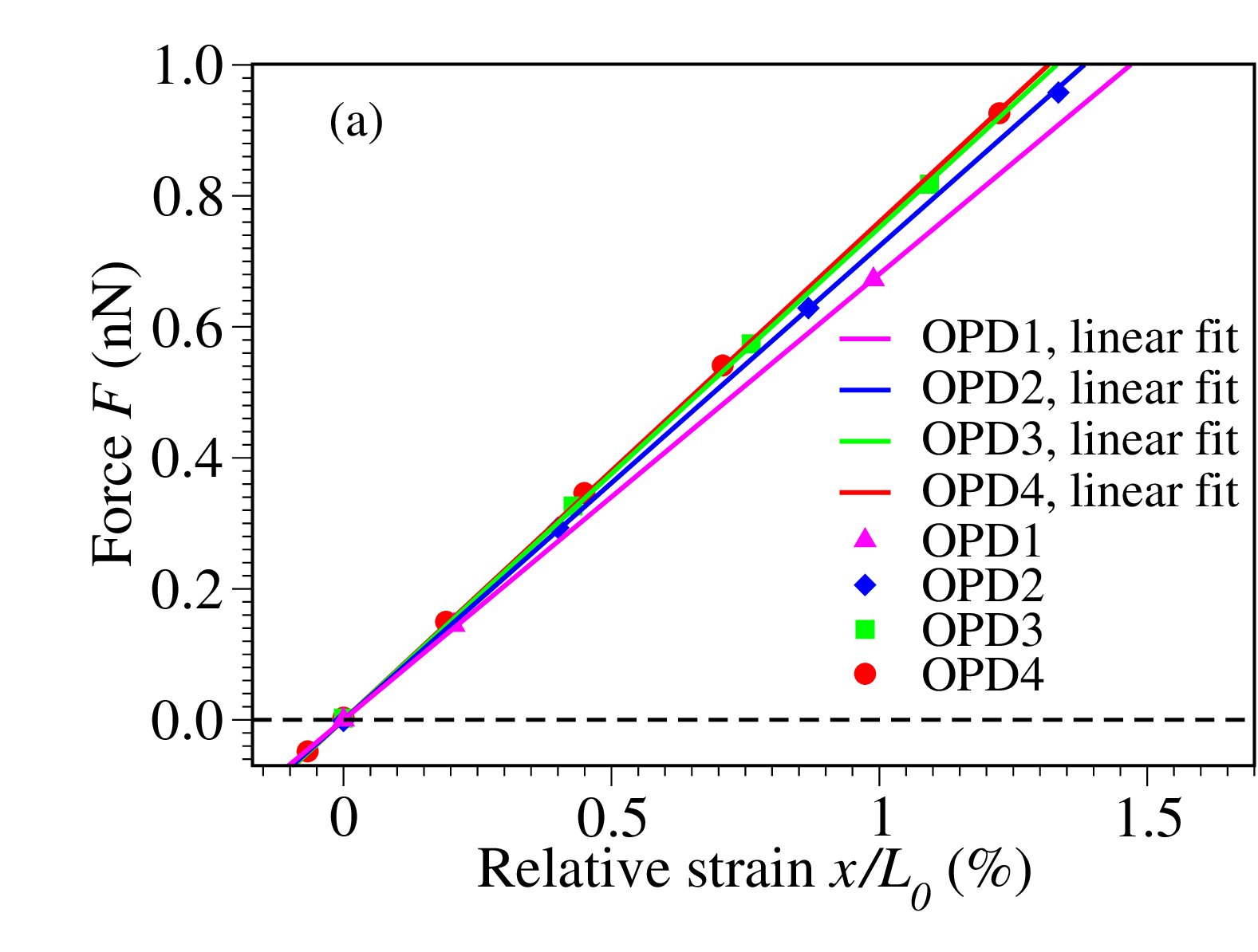}
  \includegraphics[width=0.45\textwidth]{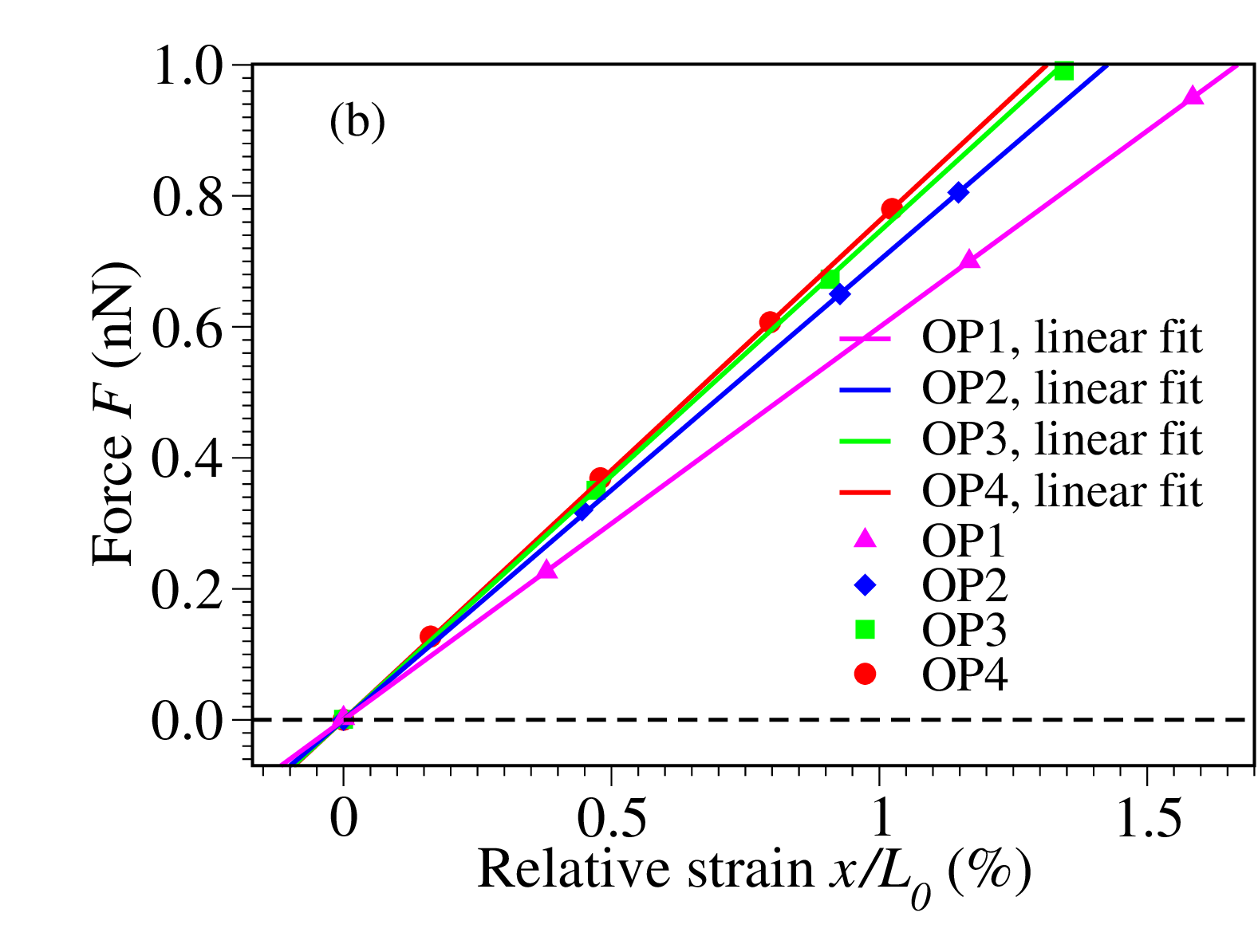}}
\caption{Results for the mechanical (here, $F>0$, tensile or compressing $F<0$) force
                 as a function of molecular length $L$: (a) OPDn and (b) OPn.} 
\label{fig:F-vs-x/L}
\end{figure*}

Results of these quantum chemical calculations at small deformations
are depicted in \figsname\ref{fig:F-vs-x} and \ref{fig:F-vs-x/L}.
The numerical values underlying these figures are collected in
Table~\ref{table:kappa-K-E}.
They reveal a linear dependence (Hooke's law) on $x$
of the force $F$ producing molecule's mechanical deformation (cf.~\figname\ref{fig:F-vs-x})
\begin{equation}
  \label{eq-k}
  F = \kappa x 
\end{equation}
Its slope provides us with the molecule's elastic (spring) constant $\kappa$. 
As visible in \figname\ref{fig:F-vs-x}a,
(straight) lines for longer OPDn molecules have larger slopes.
This is in accord with the fact that, at a given deforming force $F$,
homogeneous springs characterized by a length-independent
specific stress $K$
\begin{equation}
  \label{eq-K}
  F = K \frac{x}{L_0} ; K \equiv \kappa L_0 
\end{equation}
respond with elongations proportional to their length:
$x \propto L_0$, $\kappa \propto 1/L_0$.
In fact, OPDn molecules do not behave like homogeneous springs because
the stiffness of their constituents is different:
in the linear elastic regime presently considered (i.e.~at small $F$),
the interring \ce{C\bond{1}C} bonds are notably
stiffer than the aromatic phenyl rings, which are in turns
somewhat stiffer than the terminal \ce{C\bond{1}S} bonds. 

This results in values of the specific stress $K$ slightly decreasing from
the longer OPD4 to the shorter OPD1, as observable by inspecting the slopes of the
lines depicted in \figname\ref{fig:F-vs-x/L}a.

For comparison purposes, in \figsname\ref{fig:F-vs-x}b and \ref{fig:F-vs-x/L}b
we also present results for the elastic properties of the
parent (non-thiolated) molecules of the oligophenylene
series \ce{OPn\equiv H\bond{1}(C6H4)_n\bond{1}H} ($1\leq n \leq 4$).
The trend across the OPDn family noted above is also visible for OPn.
Notice, in particular, that OP1 ($\equiv$ benzene) has the lowest specific stress;
the stiffest interring \ce{C\bond{1}C} bond is missing there.

\begin{figure*}
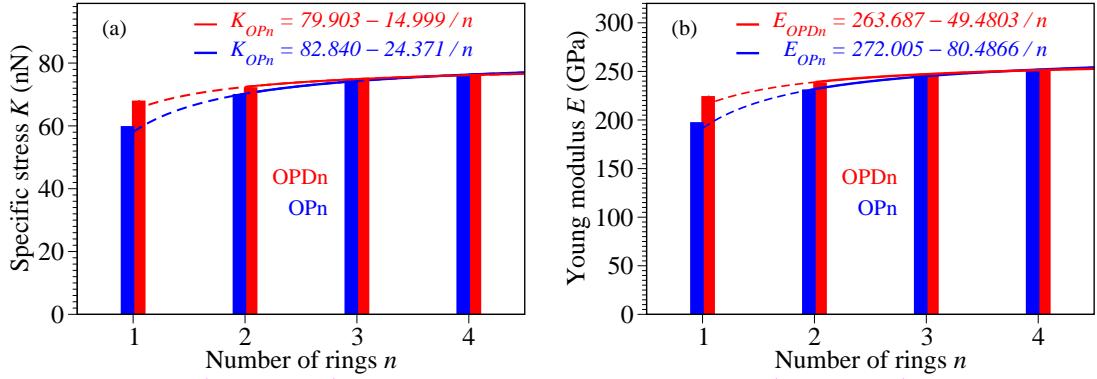

  \centerline{\includegraphics[width=0.45\textwidth]{fig_specific_stress_OPDn_OPn.eps}
    \includegraphics[width=0.45\textwidth]{fig_young_moduli_OPDn_OPn.eps}}
  \caption{Results for the specific strengths $K$ (cf.~\gl~(\ref{eq-K}))
    and Young's moduli $E$ (panels a and b, respectively)
    of oligophenylene dithiol molecules (OPDn)
    and their parent (OPn) non-thiolated species calculated microscopically
    as described in the main text and fitted as indicated in the inset.} 
\label{fig:young-modulus}
\end{figure*}

A variety of carbon-based molecular families \ce{M_n} \cite{Fan:89,Tao:07,Nijhuis:15a,Wu:16,Baldea:2019g}
are known to exhibit even-odd alternation upon adding a repeat unit (phenyl ring in our case)
$\ce{M_n} \to \ce{M_{n+1}}$.
This is a particularly meaningful question because symmetry is found to alternate across the OPDn family;
calculations indicate that the odd-number members OPD1 and OPD3 possess \ce{C_i}
symmetry while even-number members OPD2 and OPD4 possess \ce{C2} symmetry.
Nevertheless, the results depicted in \figname\ref{fig:young-modulus} reveal that  
this is not the case. The dependence on $n$ is monotonic.
Guided by the number of OPDn and OPn constituents,
we were led to consider an analytical formula used for 
the fitting curves presented in \figname\ref{fig:young-modulus},
which allows us to put the aforementioned monotonic dependence on $n$ in more quantitative terms.  

Like in cases of the rods or bars schematically depicted in
\figname\ref{fig:schematic}b,
for the characterization in terms of Young's modulus of elasticity $E$
via Hooke's law, a cross section area $A$ needs to be assigned
\begin{equation}
  \label{eq-hooke}
  F = \underbrace{E \frac{A}{L_0}}_{\kappa}  x ; \ E = \frac{\kappa L_0}{A} = \frac{K}{A} 
\end{equation}
In the specific case considered, the effective
cross section area can be expressed in terms of the surface coverage $\Sigma$
(cf.~\figname\ref{fig:schematic}b). Thanks to RBS and NRA, for OPDn SAMs used to
fabricate the CP-AFM junctions of ref.~\cite{Baldea:2017e},
this quantity is available: $\Sigma \approx 3.3\,\mbox{nm}^{-2}$ \cite{Baldea:2017e}. This yields
an average area per molecule 
\begin{equation}
  \label{eq-coverage}
  A = 1/\Sigma \approx 30\,\mbox{nm}^2
\end{equation}

The values of the Young's moduli deduced via \gls~(\ref{eq-hooke}) and (\ref{eq-coverage})
are presented in Table~\ref{table:kappa-K-E}.
They show that the SAMs used to fabricate OPDn CP-AFM junctions of ref.~\cite{Baldea:2017e}
are very stiff. They are stiffer than steel
(180\,GPa for stainless AISI 302 and 200\,GPa for structural ASTM-A36 \cite{YoungModulusSteel}).

Let us also refer to two materials used for commercial AFM cantilevers:
silicon and silicon nitride. These materials dominate the vast majority of applications \cite{Butt:05}. 
OPDn SAMs are stiffer than silicon: $E_{\ce{Si}} \approx 130 - 185$\,GPa \cite{Butt:05}.
(Precise values of real materials
depend on various factors, e.g., precise composition and crystallographic orientation).
Values for silicon nitride are in the range
$E_{\ce{Si3N4}} \approx 160 - 290$\,GPa \cite{Butt:05}.
So, using the coverages measured in ref.~\cite{Baldea:2017e}
we can also conclude that the OPDn SAMs of ref.~\cite{Baldea:2017e} can be even stiffer than silicon nitride.

Yet, this is not the whole issue. Notwithstanding that the OPDn SAMs fabricated on polycrystalline gold
in ref.~\cite{Baldea:2017e}
were found to be characterized by exceptionally small statistical variations
(hence implicitly good coverage),
the surface coverage of OPDn SAMs with herringbone (hb) arrangement
adsorbed on fcc Au (111) is higher ($\Sigma_{hb} = 4.63\,\mbox{molec/nm}^2$ \cite{Baldea:2017f})
than that of ref.~\cite{Baldea:2017e} ($\Sigma \approx 3.3\,\mbox{molec/nm}^2$).
This (hb) coverage corresponds to even higher values of $E$, which are also
included in Table~\ref{table:kappa-K-E}. Inspection of the last column of
Table~\ref{table:kappa-K-E} reveals that 
OPDn SAMs with herringbone structure are definitely stiffer than \ce{Si3N4}.
\begin{table*} 
  \caption{Elastic properties of OPDn and OPn molecules calculated microscopically as indicated in the main text: spring constants ($\kappa$), specific stresses $K$, and Young's moduli $E$.    The values of the Young's modulus $E$ correspond to two values of the surface 
    coverages $\Sigma$: (a) value measured via Rutherford backscattering (RBS) and nuclear 
    reaction analysis (NRA) for OPDn SAMs used to fabricate molecular CP-AFM junctions in 
    ref.~\cite{Baldea:2017e} ($\Sigma = 3.3\,\mbox{molec/nm}^2$)  
    and (b) value for OPDn SAMs with herringbone structure ($\Sigma = 4.63\,\mbox{molec/nm}^2$, 
    cf.~ref~\cite{Baldea:2017f} and citations therein).}
\begin{center}
  \begin{tabular*}{0.7\textwidth}{@{\extracolsep{\fill}}crrlllll}
    \hline
Molecule         &  $\kappa_{OPDn}$ & $\kappa_{OPn}$ & $ K_{OPDn} $ & $ K_{OPn} $ & $E_{OPDn}^{a} $ & $ E_{OPn}^{a} $ & $E_{OPDn}^{b} $ \\
  \hline                                                                                                                 
  OPD1       &  10.658          & 21.488         & 68.069       &  59.939     &     224.6       &    197.8        &     315.1       \\
  OPD2       &   6.755          &  9.856         & 72.331       &  70.154     &     238.7       &    231.5        &     334.9       \\
  OPD3       &   4.997          &  6.511         & 75.137       &  74.525     &     248.0       &    245.9        &     347.9       \\
  OPD4       &   3.925          &  4.829         & 76.003       &  76.176     &     250.8       &    251.4        &     351.9       \\  
    \hline
  \end{tabular*}
\label{table:kappa-K-E}
\end{center}
\end{table*}
\section{Conclusion}
\label{sec:conclusion}
To the best of author's knowledge, this is the first work reporting results for
Young's moduli of elasticity of SAMs based on benchmark species (as the case of
the presently considered family of oligophenylene dithiols OPDn)
routinely used for fabricating CP-AFM molecular junctions.
The fact that the present values $ E_{\ce{OPDn}} \approx 240 \pm 6$\,GPa
are much larger than the lower bound $ E_{\ce{OPDn}}^{exp} \approx 58$\,GPa
extracted from recent experiments \cite{Baldea:2017e} is noteworthy.

On one side, it provides a rationale for the difficulty encountered by experimentalists
to measure a reliable value of $E$;
the OPDn SAM high stiffness makes the deformation measurable for OPDn/Au in excess to that
for bare gold too small to be accurately estimated.

On the other side and more importantly from a practical standpoint,
the presently reported values of $E$ make it possible to update/refine
the numbers $N$ of OPDn molecules per junction.
At a given AFM tip and load (typically compressive $F = 1$\,nN)
which is necessarily applied to render conduction through CP-AFM junctions possible,
the contact area $A$ for stiffer SAMs
(read $E_{\ce{OPDn}} \approx 240 \pm 6$\,GPa, the presently reported values,
cf.~Table~\ref{table:kappa-K-E}) is significantly smaller than for softer  
SAMs (read $ E_{\ce{OPDn}}^{exp} \approx 58$\,GPa, as used in ref.~\cite{Baldea:2017e}).

Work is underway, but what one can state for sure is that
the ensuing values of $N$ will be even smaller than those ($N\sim 80 $)
estimated in ref.~\cite{Baldea:2017e}, which, surprisingly, already appeared much smaller
and reproducible than those (up to $N \sim 10^3$ \cite{Boer:08}) 
claimed in earlier literature \cite{Boer:08} on CP-AFM junctions.
Further, since what is directly accessible in CP-AFM experiments is a junction's transport
property (e.g., conductance $G_{junc}$, current $I_{junc}$), updated/refined $N$s' also translate
into updated/refined values of transport property per molecule ($G_1 = G_{junc}/N$, $I_1 = I_{junc}/N$).
This, in turn, makes a quantitative comparison with transport properties measured using
single-molecule (e.g., STM) testbeds more adequate.
\section*{Acknowledgments}
The author thanks Dan Frisbie and Zuoti Xie for providing him unpublished experimental data. 
Financial support from
the Deu\-tsche For\-schungs\-ge\-mein\-schaft (DFG grant BA 1799/3-2)
and computational support from the State of Baden-W\"urttemberg through bwHPC/DFG
through grants INST 40/467-1 FUGG and INST 40/575-1 FUGG (JUSTUS-2) in the initial stage of this research
is gratefully acknowledged.

\end{document}